\documentclass{nature}
\usepackage{bm}
\usepackage{hyperref}
\usepackage{amssymb}
\usepackage{epsfig}
\usepackage{color}
\usepackage{amssymb,amsmath}
\usepackage{aas_macros}
\usepackage{graphicx,epstopdf}

\newcommand{\be}{\begin{equation}}
\newcommand{\ee}{\end{equation}}
\newcommand{\ba}{\begin{eqnarray}}
\newcommand{\ea}{\end{eqnarray}}

\newcommand{\Ms}{{\ensuremath{\mathrm{M}_{\odot}}}}
\newcommand{\Rs}{{\ensuremath{\mathrm{R}_{\odot}}}}
\newcommand{\Zs}{{\ensuremath{\mathrm{Z}_{\odot}}}}
\newcommand{\Ls}{{\ensuremath{\mathrm{L}_{\odot}}}}

\title{Turbulent cold flows gave birth to the first quasars}

\author{Latif, M. A.$^{1}$, Whalen, D. J.$^{*2}$, Khochfar, S.$^3$, Herrington, N. P.$^4$ \& Woods, T. E.$^5$}

\begin{document}

\maketitle

\begin{affiliations}

\item Physics Department, College of Science, United Arab Emirates University, PO Box 15551, Al-Ain, UAE, \texttt{latifne@gmail.com}
\item Institute of Cosmology and Gravitation, Portsmouth University, Dennis Sciama Building, Portsmouth PO1 3FX, \texttt{dwhalen1999@gmail.com}
\item Institute for Astronomy, University of Edinburgh, Royal Observatory, Blackford Hill, Edinburgh EH9 3HJ, UK 
\item School of Physics and Astronomy, University of Exeter, Stocker Road, Exeter, EX4 4QL, UK
\item National Research Council of Canada, Herzberg Astronomy \& Astrophysics Research Centre, 5071 West Saanich Road, Victoria, BC V9E 2E7, Canada 

\end{affiliations}

\begin{abstract} 

How quasars powered by supermassive black holes (SMBHs) formed less than a billion years after the Big Bang is still one of the outstanding problems in astrophysics 20 years after their discovery\cite{fan03,mort11,vol12,wang21}.  Cosmological simulations suggest that rare cold flows converging on primordial haloes in low-shear environments could have created these quasars if they were 10$^4$ - 10$^5$ solar masses at birth but could not resolve their formation\cite{ten18,smidt18,hua19,zhu20}.  Semianalytical studies of the progenitor halo of a primordial quasar found that it favours the formation of such seeds but could not verify if one actually appeared\cite{lup21}. Here we show that a halo at the rare convergence of strong, cold accretion flows creates massive BH seeds without the need for UV backgrounds, supersonic streaming motions, or even atomic cooling.  Cold flows drive violent, supersonic turbulence in the halo that prevents star formation until it reaches a mass that triggers sudden, catastrophic baryon collapse that forms 31,000 and 40,000 solar-mass stars.  This simple, robust process ensures that haloes capable of forming quasars by z $>$ 6 produce massive seeds. The first quasars were thus a natural consequence of structure formation in cold dark matter cosmologies, not exotic, finely-tuned environments as previously thought\cite{an14,latif15a,hir17,tyr17,tyr21a}. 

\end{abstract}

\begin{figure}
\centering
\includegraphics[width=\textwidth]{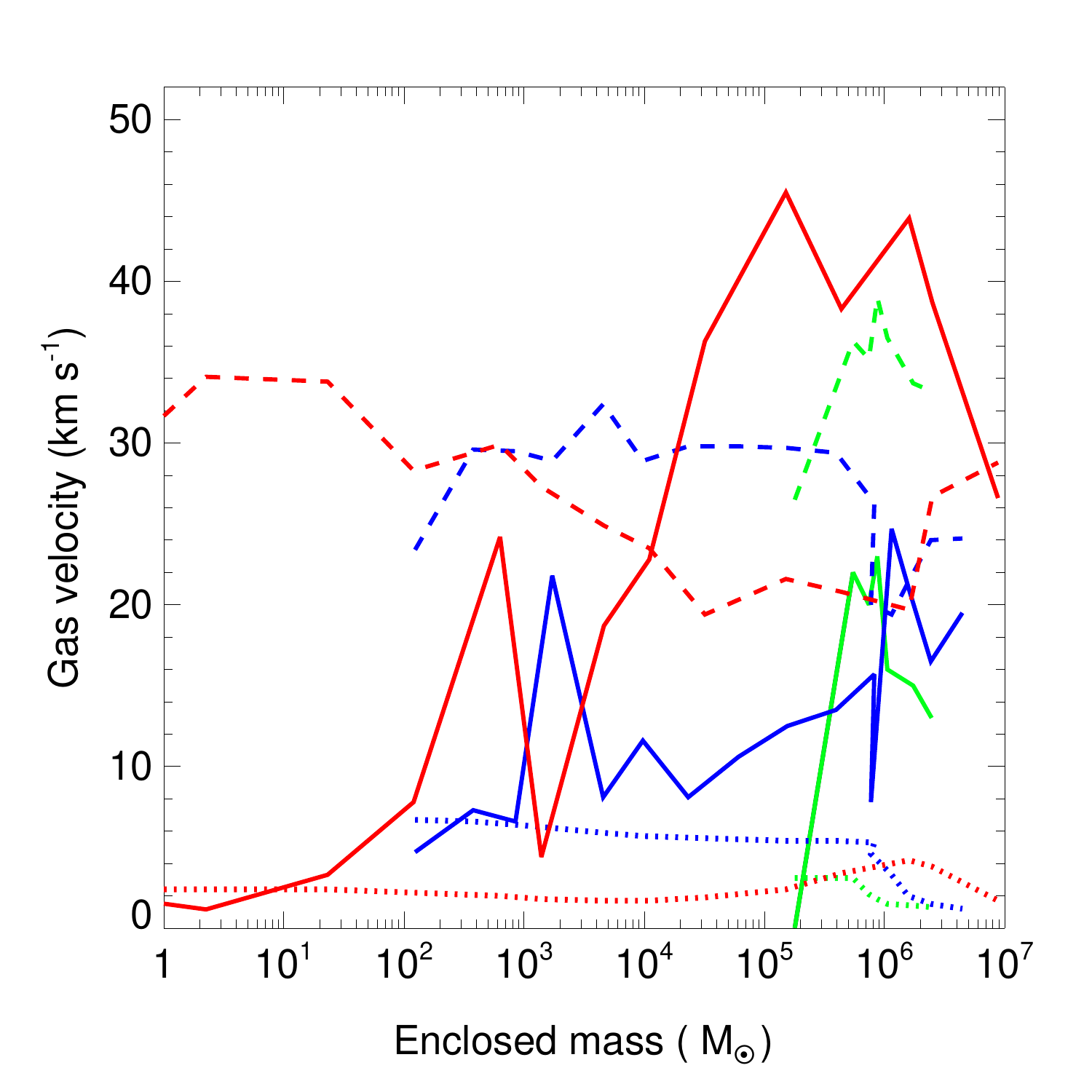}
\caption{{\bf Spherically-averaged velocity profiles of the halo.}  The infall velocities (solid), turbulent velocities (dashed) and sound speeds (dotted) are at just before collapse (z $=$ 29, green), the onset of collapse (z $=$ 25.74, blue), and well into collapse (z $=$ 25.73, red).  Enclosed gas masses of 10$^4$ \Ms\ and 10$^6$ \Ms\ correspond to radii of about 1 pc and 100 pc, respectively.}
\label{fig:vturb}
\end{figure}

We investigated SMBH seed formation during the earliest stages of collapse of the progenitor halo of a z $>$ 6 quasar with the Enzo cosmology code.  This halo grows to 1.4 $\times$ 10$^{12}$ \Ms\ (\Ms, solar mass) by z $=$ 6 mostly by cold flow accretion\cite{yu14,lup19,vgf21} rather than by major mergers with other gas-rich haloes\cite{li07}, and it is not in close proximity to any external sources of Lyman-Werner (LW) UV flux or in any LW backgrounds when it first collapses.  Our simulation resolves the halo when it  reaches a mass of 3.9 $\times$ 10$^5$ \Ms\ at z $=$ 35, the approximate minimum mass at which it could form Population III (Pop III) stars.  However, highly supersonic turbulence driven by the cold flows from large scales prevent gas in the halo from collapsing into stars at this mass.  As shown at z $=$ 29 in Figure~\ref{fig:vturb}, turbulent velocities are 30 - 40 km s$^{-1}$, a factor of a few higher than the infall velocities and more than an order of magnitude greater than the sound speed of the gas.  

\begin{figure*}
\begin{center}
\includegraphics[width=\textwidth]{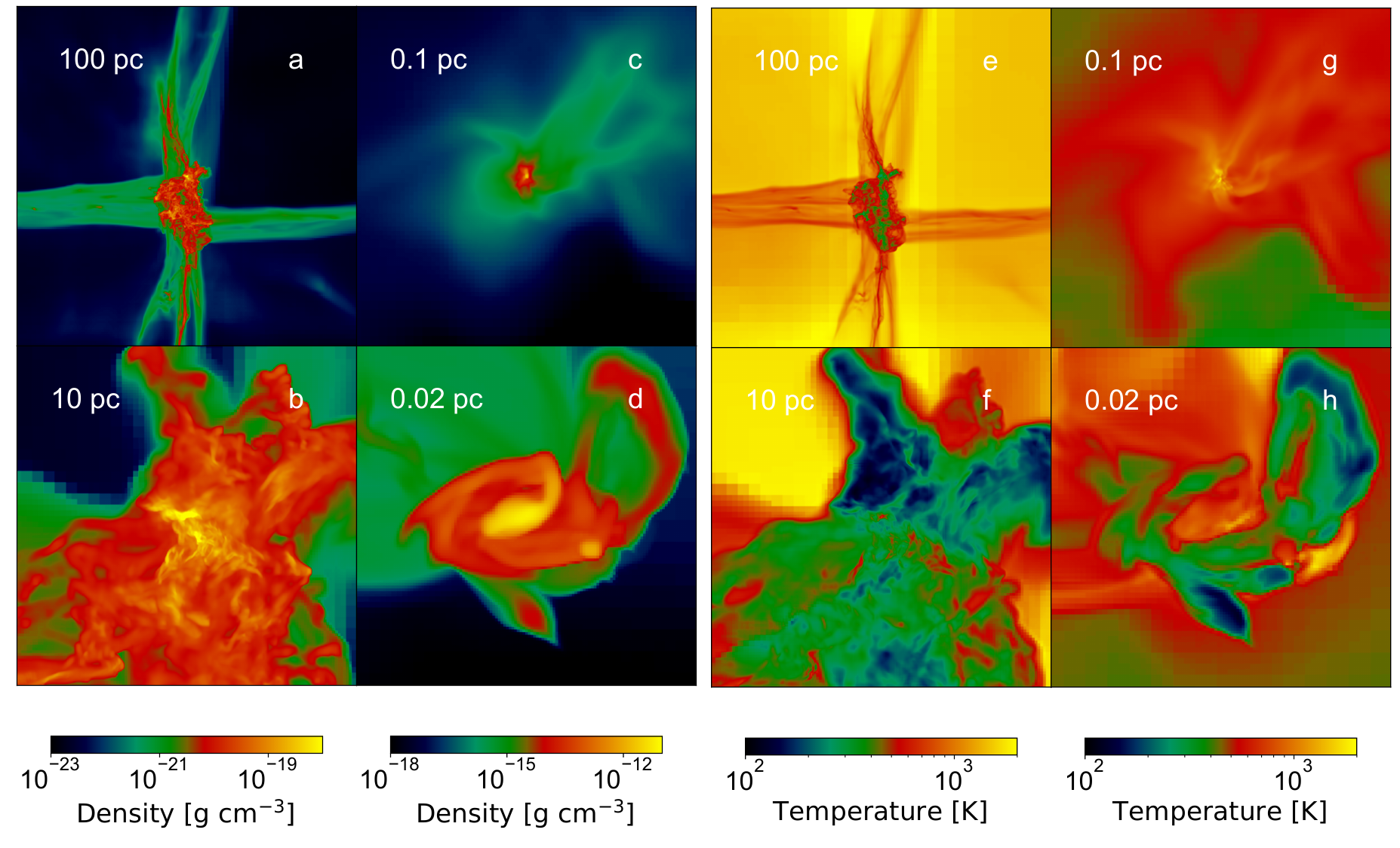} 
\end{center}
\caption{{\bf Catastrophic baryon collapse in the halo at z $=$ 25.7.}  Densities are in a - d and temperatures are in e - h where a and e are at the onset of collapse, b and f are at 4.5 kyr,  c and g are at 5.5 kyr, and d and h are at 11 kyr.}
\label{fig:halo}
\end{figure*}

Over the next 45 Myr the halo grows by two orders of magnitude in mass, to 4 $\times$ 10$^7$ \Ms\ by z $=$ 25.7 at an average rate of $\sim$ 0.9 \Ms\ yr$^{-1}$.  Although this is roughly four times the mass at which atomic cooling normally begins\cite{wise19} ($\rm 2.2 \times 10^7~\Ms ~[(1+z)/16]^{-3/2}$), the gas does not collapse by Ly$\alpha$ emission because temperatures in the halo never exceed 2,000 K.  Collapse is instead triggered when infall velocities finally exceed turbulent velocities because the halo has grown so much in mass, as shown in Figure~\ref{fig:vturb}, and gravitational compression overcomes turbulent pressure support.  We show density and temperature projections of the halo at the onset of collapse in Figure~\ref{fig:halo}.  The four cold accretion streams seen in the 100 pc image carry gas at densities of 10$^3$ - 10$^4$ cm$^{-3}$ deep into the halo and create the highly turbulent core visible in the 10 pc panel.  Energy transfer from scales of tens of kpc down to pc scales drives this turbulence.  A dense clump appears in the core 5 kyr after the onset of collapse, as shown in the 0.1 pc image.  It forms a small self-gravitating disk with two prominent spiral arms and a mass of about 1,000 \Ms, as shown in the 0.02 pc panel.  Analysis indicates that it is stable to fragmentation, and it rotates at nearly the Keplerian velocity with influx rates of about 1 \Ms\ yr$^{-1}$.   However, before this clump can form a star it becomes part of a much larger clump that forms a supermassive star later on (C2, as discussed below).

\begin{figure}
\centering
\includegraphics[width=\textwidth]{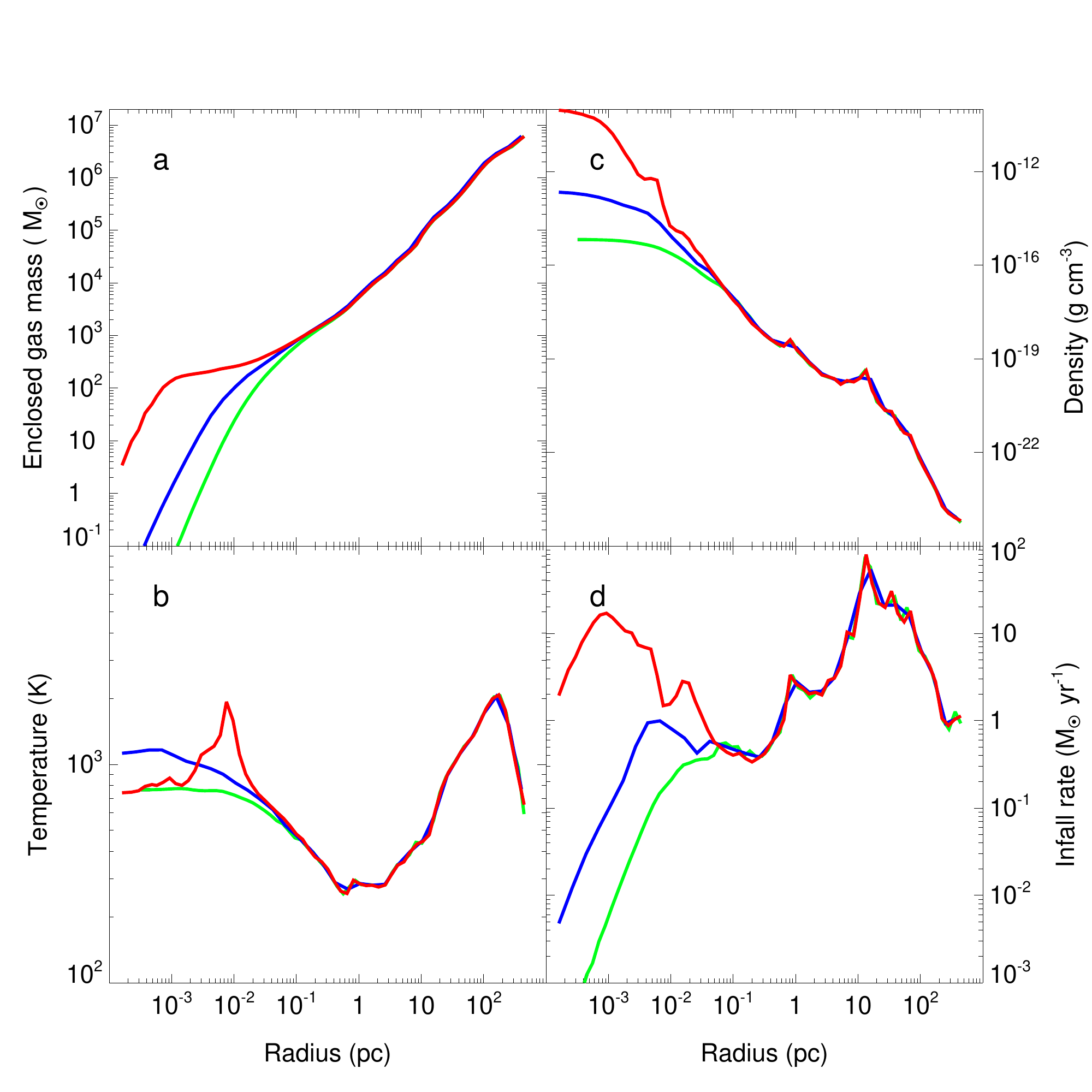}
\caption{{\bf Spherically-averaged gas profiles of the halo.} The enclosed masses (a), temperatures (b), densities (c), and inflow rates (d) are for the onset of collapse (green), 5 kyr (blue), and 11 kyr (red).}
\label{fig:cprof}
\end{figure}

Unlike collapse in normal Pop III star-forming haloes, which creates quasi-static settling flows\cite{bl04c}, or in atomically-cooled haloes in which there is at most transonic turbulence, collapse in this halo proceeds on two scales via two, distinct physical processes.  As shown in Figure~\ref{fig:cprof}, it begins on scales of 20 - 200 pc when cold flows finally overcome the turbulent pressure they originally created and settle onto the core of the halo at influx rates of up to 40 \Ms\ yr$^{-1}$.  The pileup of gas at the outer layers of the core is visible in the peaks in the density and infall velocity profiles at 20 pc.  Collapse on these scales is driven mostly by dynamics.  Infall at larger scales compresses gas to densities of 10$^{-16}$ g cm$^{-3}$ at radii below 0.01 pc, activating rapid H$_2$ formation via three-body reactions that almost fully molecularise the core by 11 kyr as shown in Extended Data Figure~\ref{fig:JM}.  Rapid cooling by these large H$_2$ mass fractions causes a second stage of collapse at radii below 0.1 pc, with inflow rates of up to 5 \Ms\ yr$^{-1}$ even down to 20 AU, the smallest scales resolved by our simulation at this stage.  Densities at the center of the halo rise by four orders of magnitude to $\sim$ 10$^{-11}$ g cm$^{-3}$ by 11 kyr.  Gravitational compression briefly raises temperatures at the center of the halo from 750 K to 1,200 K before rapid H$_2$ cooling drops them back down to 700 K.  The dramatic rise in density at the center of the halo causes gas to pile up at radii of 0.01 pc, where a shock forms and heats the gas to 1,700 K at 11 kyr (this shock is also visible in the dip in infall rates at the same radius).  

\begin{figure}
\centering
\includegraphics[width=\textwidth]{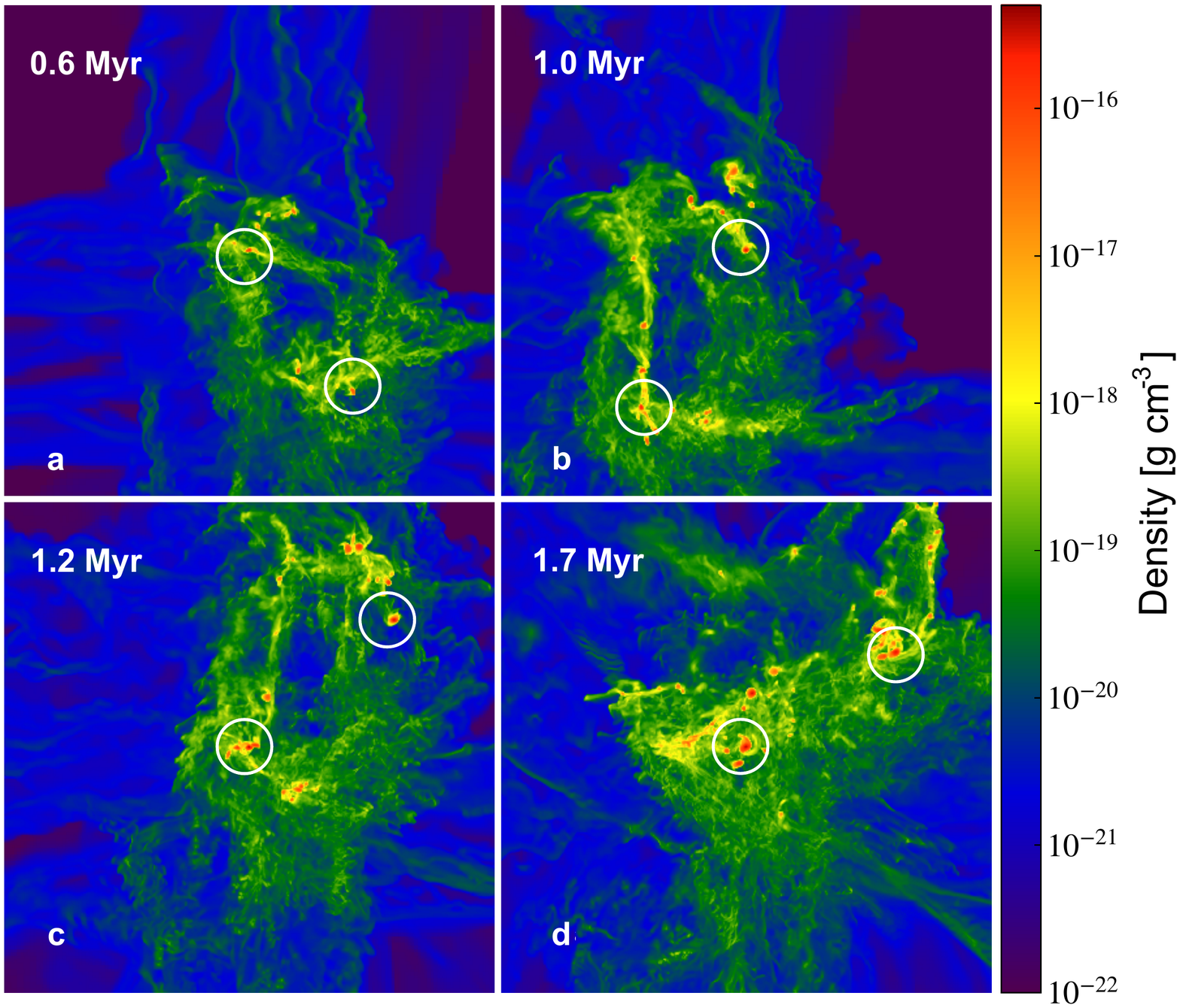}
\caption{{\bf Evolution of the turbulent core.}  C1 and C2 are marked by the white circles and the images are 25 pc on a side.  The times in a - d are 0.6 Myr - 1.7 Myr after collapse.}
\label{fig:core}
\end{figure}

While resolutions of 20 AU capture the initial stages of collapse on all relevant scales they are not practical for following massive seed formation for 1 - 2 Myr because of short Courant times. We halted the simulation, restarted it from a slightly higher redshift (z $=$ 25.73), when the halo was only resolved by 18 levels of refinement (0.01 pc), and then evolved the halo at this new limit of refinement for 1.6 Myr to determine if it produces a supermassive star (SMS).  As shown in Figure~\ref{fig:core}, turbulence prevents the formation of any coherent structures in the halo like the large accretion disks that form during the collapse of atomically-cooled haloes\cite{ln06,rh09b}.  The core of the halo instead develops a wispy, filamentary structure that is interspersed with dense clumps.  The two most massive ones, C1 and C2, appear in the first 100 kyr.  C1 initially grows mainly through accretion but merges with several other clumps 200 kyr later.  As shown in Extended Data Figure~\ref{fig:cmass}, these collisions raise its mass to 10$^4$ \Ms\ over 400 kyr and it then doubles in mass over the next 600 kyr.  Its growth later falls off because it migrates into regions with lower gas densities.  C1 reaches a mass of 2.6 $\times$ 10$^4$ \Ms\ just before merging with another cluster of clumps at the end of the run.

C2 grows to 1,000 \Ms\ in 100 kyr before multiple encounters with dense filaments increase its mass by a factor of four over the next 100 kyr.  It then grows to 8,000 \Ms\ in about 900 kyr primarily by accretion.  The four cold accretion streams impart angular momentum to the core and create the transient annular structure shown in Fig.~\ref{fig:core} that persists for about a Myr.  An elongated dense filament forms in the annulus and fragments into multiple clumps.  C2 merges with a few of these clumps and  grows by a factor of five in mass in about 200 kyr.  The other clumps orbit C2, which briefly forms spiral arms due to tidal interactions with them, but most of them eventually merge with C2 except for one that is ejected from the cluster and two that remain in its orbit.  C2's mass at the end of the simulation is 4.1 $\times$ 10$^4$ \Ms. 
 
C1 and C2 collapse before any stars form in the halo, at infall rates of 0.01 - 0.55 \Ms\ yr$^{-1}$ (left panel of Extended Data Figure~\ref{fig:mdot}).  These rates are on par with those in atomically-cooled haloes that form SMSs in 1 - 2 Myr\cite{pat21a}.  We modeled the evolution of the stars that form in these clumps with the MESA code and show their Hertzsprung-Russell diagrams in the right panel of Extended Data Figure~\ref{fig:mdot}.  Both initially evolve along the Hayashi track as cool, red hypergiants because of H$^-$ opacity in their outer layers, which causes them to expand and remain at approximately 5,500 K as the stars grow in mass.  However, during periods of quiescence when there is little or no accretion at later times, the stars can contract onto hotter and bluer tracks and both eventually reach temperatures of 100,000 - 120,000 K.  SMSs 1 and 2 collapse at masses of 31,316 \Ms\ and 39,765 \Ms, respectively.  Although both stars are millions of times more luminous than the Sun, neither will be visible to the James Webb Space Telescope\cite{sur18a} but their BHs could be detected at z $\gtrsim$ 20 after growing to 10$^5$ \Ms\cite{wet20b}.  Since both stars become blue and hot later in their lives, their ionizing UV flux could reduce their accretion rates\cite{latif21a} so these masses should be taken to be upper limits.  Nevertheless, it is clear that dynamically-driven collapse creates two DCBHs that are viable seeds for a quasar forming by z $\sim$ 6 - 7. 

Our simulations show that the number of quasars at z $>$ 6 does not depend on LW backgrounds so it is not subject to the large uncertainties in the critical flux, J$_{\mathrm{crit}}$, required to form DCBHs in atomically cooled haloes, which can vary several orders of magnitude\cite{agarw15,rosa17}.  The number of haloes that form quasars by z $\sim$ 6 in large-scale numerical simulations\cite{dm17} is 1 - 2 $\times$ 10$^{-8}$ cMpc$^{-3}$, within an order of magnitude of the number of quasars observed at z $>$ 6, about 10$^{-9}$ cMpc$^{-3}$ or 1 cGpc$^{-3}$.  Delayed collapse due to turbulence driven by cold flows therefore only has to occur in about 10\% of these haloes to account for the number of quasars at z $>$ 6 observed to date.  Because the topology of cold flows in the primordial Universe governed the numbers of these unusual reservoirs, and they could form their own massive seeds regardless of proximity to LW sources, the demographics of the first quasars may ultimately have been determined by the structure of the cosmic web.



\newpage

\noindent {\bf Methods}    

Enzo\cite{enzo} is an adaptive mesh refinement (AMR) cosmology code with an $N-$body particle-mesh scheme for evolving dark matter (DM)\cite{efs85,couch91} that is self-consistently coupled to hydrodynamics, nonequilibrium primordial gas chemistry\cite{anet97}, and a multigrid Poisson solver for self-gravity.  Our simulations utilise the piecewise parabolic method for hydrodynamics\cite{wc84,bryan95} and the HLLC scheme for enhanced stability with strong shocks and rarefaction waves\cite{toro94}. The reaction network follows the thermal and chemical evolution of the gas by evolving nine primordial species (H, He, e$^-$, H$^+$, He$^+$, He$^{2+}$, H$^-$, H$_2$ and H$_2^+$) with updates to the gas energy that include collisional excitational and ionizational cooling by H and He, recombination cooling, H$_2$ cooling, bremsstrahlung cooling, and inverse Compton cooling by the cosmic microwave background (CMB).  Our H$_2$ cooling rates\cite{ga08} are valid at low densities and high densities, at which local thermodynamic equilibrium (LTE) effects become important.  They also include the transition from optically-thin to optically-thick cooling at densities when H$_2$ becomes opaque to its own cooling lines\cite{ra04}.  Our chemistry model omits collisional ionizations of H by H and contributions to the H$_2$ collisional dissociation rate by dissociative tunnelling, which are important when strong LW fields are present\cite{g15a,g15b}.  However, since there are no LW fluxes in our simulation and the temperatures at which these two reactions become significant are 8000 K and 2000 - 4500 K, respectively (which are greater than those in our halo), their omission has little impact on our results.  The choice of chemistry network, reaction coefficients and update scheme can lead to some variations in the thermal evolution of the gas.

\subsection{Halo selection and simulation setup.}

We initialise our simulations at z $=$ 150 with cosmological initial conditions generated with MUSIC\cite{hahn11} using the second-year \textit{Planck}\cite{planck2} best fit lowP + lensing + BAO + JLA + H$_0$ cosmological parameters:  $\Omega_{\mathrm{M}}=$ 0.3089, $\Omega_{\Lambda}=$ 0.691, $\Omega_{\mathrm{b}} = $ 0.0486, h $=$ 0.677, $\sigma_8 = $ 0.816, and n $=$ 0.967.  The simulation was performed in a 37.5 Mpc box.  Given that there are only a few dozen haloes per Gpc$^{3}$ at the nexus of converging cold flows that sustain rapid early quasar growth in large-scale cosmological simulations, no single 37.5 Mpc box at random would be expected to enclose one\cite{kh12}.  We therefore evolved dozens of such boxes\cite{tss08} with different random seeds at a low top-grid resolution of $\rm 256^3$ and DM mass resolution of 10$^8$ \Ms\ until we found a halo that reached 1.4 $\times$ 10$^{12}$ \Ms\ by z $=$ 6 primarily by accretion rather than major mergers (i.e., collisions with other haloes with masses greater than 10\% of the mass of the host halo).  

To capture the assembly history of this halo in greater detail the simulation was rerun with a DM particle mass of 2.3 $\times$ 10$^5$ \Ms\ and outputs at 10 Myr intervals.  A halo catalogue for each output was then computed with Rockstar\cite{rockstar} and the merger history of the halo was constructed from the catalogues with the consistent-tree algorithm.  The mass of the halo over time is shown in Extended Data Figure~\ref{fig:hmass}.  Analysis shows that the halo grows mainly by accretion and minor mergers with mass ratios below 1:10, having experienced only one major merger with a mass ratio of 1:3 at z $=$ 11.3 and two mergers with mass ratios of 1:10 at z $=$ 12.

We then identified the DM particles in the Lagrangian volume of the halo at z $=$ 6 with Rockstar, traced their masses and positions back to the initialization redshift, added one level of refinement in the region that encloses these particles (twice the Lagrange radius of the halo), and reran the simulation down to z $=$ 6.  We again identified the DM particles of the now better-resolved halo, traced their masses and positions back to z $=$ 150, applied another level of refinement to the volume enclosing this new particle set, and evolved  the simulation to the final redshift.  This procedure\cite{mrp} was repeated until it produced five nested grids with a maximum DM resolution of 3,636 \Ms, an effective initial spatial resolution of 8,192$^3$, and 975 million DM particles.  The minimum DM particle mass allows us to resolve the halo at the lowest mass at which it could have formed normal Pop III stars, a few 10$^5$ \Ms.   

Our simulation is partitioned into two stages.  In the first, we evolved the halo from z = 150 with DM and gas down to the onset of collapse at z = 25.7 while allowing up to 25 additional levels of refinement in the innermost nested grid for a maximum resolution of 20 AU.  Our refinement criteria are based on baryonic overdensity, Jeans refinement and the DM mass resolution.  A cell is flagged for refinement if it exceeds 2.5 times the cosmic mean density or a DM particle density of 0.0391 $\rho_{\rm{DM}}r^{\ell \alpha}$, where $\rho_{\rm{DM}}$ is the dark matter density, r $=$ 2 is the refinement factor, $\ell$ is the refinement level, and $\alpha = -0.2$, which enforces super-Lagrangian refinement.  The Jeans length in the simulations was always covered by at least 64 cells in order to resolve turbulent eddies in the flows\cite{fed11,latif13c}.  We also used the must-refine particle flag, which guarantees that DM particles are refined to at least their generation level. 

When the simulation reached 25 levels of refinement, we halted it and imposed a pressure floor on the highest level to prevent collapse to even smaller scales.  This was done by artificially setting temperatures in the cells to ensure they are Jeans stable and cannot fragment.  We then evolved the halo with this floor for 20 kyr to follow its collapse at a resolution of 20 AU.   In the second stage of the run we restarted the simulation from a slightly earlier time (z $=$ 25.73), when it was only at a maximum of 18 levels of refinement (a resolution of $\sim$ 0.01 pc), and limited the AMR to 18 levels thereafter to be able to evolve the halo for the times required for the two SMSs to form and collapse, about 1.7 Myr.  We set a new pressure floor on the 18th level so gas would not collapse below our resolution limit. 

\subsection{H$_2$ Cooling.}

H$_2$ cooling can dissipate virial shock heating and cool a halo of this mass to less than 2,000 K because its H$_2$ mass fractions remain high in the absence of nearby sources of LW photons, and at 1,000 - 2,000 K the H$_2$ cooling rates can be 10 - 100 times higher than those in normal Pop III star-forming haloes.  Densities at the virial radius of this halo are also several times those of other haloes of comparable mass and also enhance its cooling.  A few cells reach 8,000 - 10,000 K and cool via Ly$\alpha$ emission but their total mass is small and they have no effect on the collapse.  As shown in Extended Data Figure~\ref{fig:JM}, H$_2$ mass fractions vary from 0.1 - 0.8 at radii less than 0.01 pc to a few 10$^{-3}$ out to the boundary of the turbulent core at 10 - 20 pc and then down to 10$^{-4}$ at 100 pc.  H$_2$ cooling generally keeps temperatures in the halo below 1,000 K, the 2,000 K peak at 100 pc is due to virial heating by the accretion shock.  The densities exhibit significant departures from classic $r^{-2}$ isothermal collapse profiles because temperatures vary by an order of magnitude in the halo and collapse proceeds in two stages at two different scales.  The gas enclosed by the halo is 4 $\times$ 10$^6$ \Ms, two orders of magnitude higher than in normal Pop III star-forming haloes and a factor of a few higher than conventional direct-collapse black hole (DCBH) host haloes.  

\subsection{Pop III Star Formation.}

Although we did not have star or sink particle formation turned on, no normal Pop III stars formed in the halo during the first stage of the simulation because (extremely) large turbulent velocities prevented the collapse of the gas, even though it was cooled by H$_2$.  In molecular clouds today, supersonic turbulent shocks can create density fluctuations that can collapse to stars via efficient cooling by dust, metals and molecules\cite{hc08,fed18}.  Such shocks produce similar structures in our simulations but they do not form stars because H$_2$ cooling is much less efficient.  Gas in the halo did not reach high enough densities to trigger Pop III star formation before global collapse in the halo took place.

No Pop III stars formed during the collapse of the halo in the second stage of the run before the two supermassive stars appeared because supersonic turbulence driven by the cold flows modified the Jeans mass of the gas:
\begin{equation}
M_{\rm Jeans} \, = \, \frac{\pi}{6} \frac{v_{\rm eff}^3}{G^{3/2} \rho^{1/2}} ,
\end{equation}
where 
\begin{equation}
v_{\rm eff} \, = \, \sqrt{c_{\rm s}^2 + v_{\rm t}^2}.
\end{equation}
Here, v$_{\rm t}$ is the turbulent velocity\cite{stacy11b} and c$_{\rm s}$ is the sound speed. After the onset of collapse, densities in the core rise to over 1 $\times$ 10$^{-16}$ g cm$^{-3}$ and v$_{\rm t}$ and c$_{\rm s} $ are $\sim$ 20 km s$^{-1}$ and 2 km s$^{-1}$, respectively, yielding an effective Jeans mass of about 10$^4$ \Ms.  Consequently, in spite of H$_2$ cooling, the gas could only fracture on large mass scales and no normal Pop III stars formed, just the two massive clumps that created the 31,000 \Ms\ and 39,000 \Ms\ stars.  No star or sink formation was turned on during the second stage of the simulation either, so to identify potential sites of SMS formation we used the clump finder in the yt analysis package\cite{yt} to search for objects that are gravitationally bound, are resolved by a minimum of 20 cells, and have threshold densities $\geq$ 10$^{-17}$ g cm$^{-3}$.  C1 and C2 were the first self-gravitating clumps to form SMSs after the halo began to collapse, no other clumps produced stars before the SMSs became DCBHs.

The pressure floors we imposed on the highest level of refinement at the end of stage 1 and during stage 2 suppress fragmentation on scales of at most 10 times their resolution limits (200 AU and 0.1 pc, respectively\cite{met01}).  Fragmentation can occur in accretion disks on scales below our resolution\cite{get11,bec15,bec18}, but these low-mass clumps are expected to migrate inwards and merge with the central object on timescales that are much shorter than its Kelvin-Helmholz contraction times onto the main sequence.  This has been found to be true in 3D simulations\cite{hos16} and analytical studies of Pop III disk fragmentation\cite{ls15,ih14}.

We confirmed that no Pop III stars formed in the halo by turning on sink particles instead of pressure floors in stage 1 and for the first Myr of stage 2.  Sinks were created if five criteria were met:
\begin{enumerate}
\item{The grid cell is at the maximum refinement level.}
\item{The gas density in the cell is higher than the Jeans density.}
\item{The flow is convergent ($\nabla \cdot \mathbf{v} <$ 0).}
\item{The cooling time is shorter than the free-fall time.}
\item{The cell is at a local minimum of the gravitational potential.}
\end{enumerate}
Besides accreting gas, these sinks, once created, also act as star particles so they can be  sources of ionizing UV radiation.  This algorithm has been tested and used in a number of cosmological simulations\cite{rd18b,ret20,latif21a,latif21b} and is similar to those used in studies of star formation in molecular clouds today\cite{krum04,fed10}. No sink particles formed during either stage of the run, confirming that no Pop III stars formed before the two SMSs.  This outcome is consistent with previous work that found essentially no differences when using sink particles or pressure floors in simulations of DCBH formation\cite{latif13d}.  We show images of the halo 0.6 Myr and 1.0 Myr after the onset of collapse in our sink particle run in Extended Data Figure~\ref{fig:core2}.  They exhibit morphologies that are quite similar to those in Figure~\ref{fig:core}.  As shown in Extended Data Figure~\ref{fig:cmass}, C1 and C2 have similar masses in the two runs and were the first to collapse and form SMSs in them.  They also exhibit central infall rates that are similar to those in the original run, showing that pressure floors had little effect on the evolution of the core of the halo.

Our results are also consistent with a recent simulation in which supersonic gas streaming motions relative to the DM potential of a primordial halo\cite{th10} modify the Jeans mass of the gas and prevent star formation in the halo until it grows to over 10$^7$ \Ms\ and creates a SMS\cite{hir17}.  Turbulence due to 3$\sigma$ fluctuation streaming velocities suppressed collapse until the halo grew to 2 $\times$ 10$^7$ \Ms\ and formed the star via H2 and atomic cooling.  As in our simulation, no Pop III stars formed before the halo reached this mass.  Although the origin of the supersonic turbulence in our study is quite different, its suppression of star formation prior to collapse is similar.

Pop III stars are expected to eventually appear in the halo because X-rays from the DCBH can catalyze H$_2$ formation and trigger their formation\cite{mba03,aycin14,aycin20}, and they play an important role in regulating the growth of the BH at later times\cite{smidt18}. But even if normal Pop III stars had formed before the two SMSs they probably would not have been able to interfere with their growth.  Recent radiation hydrodynamical simulations indicate that such stars would have had to be at least 1,000 \Ms\ before their ionizing UV flux could disrupt massive inflows in their vicinity in similar environments\cite{latif21a}.

H$_2$ cooling has previously been shown to play a role in SMS formation when large reservoirs of gas and high halo temperatures are present.  Only the most extreme LW backgrounds destroy all the H$_2$ in a halo because it can self-shield in the high densities at the center. Not even energy injection by violent mergers destroy it all, even in moderate LW backgrounds, and such haloes have still been found to form SMSs\cite{wise19,ret20}.  A 10$^4$ \Ms\ star even formed in the presence of metals in one simulation\cite{chon20a}, in which Z $<$ 0.001 \Zs\ (\Zs, solar metallicity).  These studies show that any rapid dynamical collapse can produce SMSs, regardless of origin\cite{lv15,rd18b}.

\subsection{Jeans masses and H$_2$ mass fractions at the onset of collapse.}

In Extended Data Figure~\ref{fig:JM} we plot the ratio of the enclosed gas mass to the Jeans mass, $\rm M_{en}/M_{J}$, at the onset of dynamically-driven collapse, at 5 kyr, and at 11 kyr.  Wherever this ratio exceeds one the gas is prone to collapse.  The three plots show that the gas in the halo becomes increasingly unstable to collapse over time, with the Jeans instability operating at the smallest radii resolved in the simulation by just 11 kyr.  The peaks in the ratio indicate that gravitational instabilities are present at multiple scales and that almost all the gas in the halo is destined to collapse.  This trend is also visible in the infall rates, $\rm 4 \pi R^2 \rho v $, plotted as a function of enclosed mass at the same times in Extended Data Figure~\ref{fig:JM}.  At 5 kyr collapse rates at the center of the halo are modest, a few 0.01 \Ms\ yr$^{-1}$, but rise to nearly 1 \Ms\ yr$^{-1}$ by 11 kyr.  

\subsection{Dynamical heating.}

Recent studies have examined the formation of SMSs in atomically-cooled haloes in marginal LW backgrounds via dynamical heating due to rapid growth driven by major mergers\cite{wise19,ret20}.  In the first study (W19), a halo reaches 2 $\times$ 10$^7$ \Ms\ by $z \sim$ 16.5, having grown by a factor of 30 in mass in just 30 Myr from $z = $ 21 - 19 because of a major merger.  It was exposed to a LW flux of 3 $J_{21}$ and collapsed at a mass of $\sim$ 2 $\times$ 10$^6$ \Ms.  Dynamical heating of the gas in the halo due to rapid inflows suppressed normal Pop III star formation even though the LW background was small.

Although our halo also hosts rapid infall, dynamical heating plays little or no role in the prevention of normal Pop III SF in it prior to catastrophic collapse.  While its average growth rate is larger than in W19, below 3 $\times$ 10$^6$ \Ms\ its instantaneous growth rate is a few 10$^5$ \Ms\ per unit redshift, an order of magnitude smaller than in W19.  The maximum growth rate of our halo is still a factor of 2 - 3 times below that in W19 and peaks when the halo has already reached 10$^7$ \Ms.  Typical H$_2$ mass fractions in W19 are 10$^{-5}$ - 10$^{-6}$, more than an order of magnitude lower than those in our halo, $\sim$ 10$^{-4}$.  At these H$_2$ fractions the growth rate of our halo would have to exceed those in W19 by an order of magnitude for dynamical heating to be effective (see their Extended Data Figure 1) so it has little or no impact on collapse.  We also note that turbulent and radial infall velocities in W19 are about a factor of ten lower than in our halo so turbulence has little effect on the collapse of their halos.  The conditions leading to the formation of SMSs in our halo are thus quite distinct from those in W19.  

\subsection{SMS evolution and DCBH birth.}

We use the Modules for Experiments in Stellar Astrophysics (MESA) code\cite{paxt11} to follow the growth of the SMSs in the halo.  MESA is a one-dimensional lagrangian stellar evolution and hydrodynamics code with nuclear burning that is implicitly coupled to updates of the equations of stellar structure\cite{paxt13,paxt18}.  We initialise the stars as 10 \Ms, fully convective (n $=$ 1.5) polytropes with hydrogen and helium mass fractions of 0.76 and 0.24, respectively, and central densities $\rho_{\mathrm{c}} =$ 7.6 $\times$10$^{-6}$ g cm$^{-3}$ and temperatures T$_{\mathrm{c}} =$ 9.0 $\times$ 10$^5$ K.  

MESA adaptively rezones the stars over their evolution. The mesh is refined when gradients in pressure, temperature and $^4$He abundances between cells exceed preset limits: $\delta \rm \log P/P >$ 1/30, $\delta \rm \log T/T >$ 1/80, and $\delta \log(\chi + 0.01) >$ 1/20 $\log \chi$, where $\chi$ is the $^4$He mass fraction.  This approach typically partitions the stars into 750 - 1,200 mass zones, with larger numbers of zones being allocated to the center of the star to resolve nuclear burning and convective processes.  We use the 21-isotope \texttt{APPROX21} nuclear reaction network, which includes the p-p chain, CNO cycle, helium burning, and the CO and alpha chains.    

We apply the Ledoux criterion for convection and use Henyey mixing-length theory\cite{hen65}.  The equation of state (EOS) in our models is a composite of several datasets, including the OPAL/SCVH tables\cite{opal3,sau95}, which are used at lower densities and temperatures like the outer regions of the star and its atmosphere, and the \texttt{HELM} and \texttt{PC} EOSs,\cite{ts00,pot10} which are used in the core of the star.  Above 10$^4$ \Ms\ stars can encounter the Chandrasekhar-Feynman general relativistic (GR) instability, in which thermal photons in their cores reach densities at which they can enhance gravity via the stress-energy tensor and cause the core to collapse.\cite{chandra64}  This effect can be approximated by the post-Newtonian Tolman-Oppenheimer-Volkoff (TOV) correction to the equation of hydrostatic equilibrium,
\begin{equation}
       \frac{dP}{dr} = -\frac{Gm\rho}{r^2},
       \label{eqn:HE}
\vspace{0.1in}
\end{equation}
by replacing the gravitational constant G with G$_{\rm rel}$,
\begin{equation}
	G_{\textrm{rel}} = G\left(1 + \frac{P}{\rho c^2} + \frac{4\pi Pr^3}{m_r c^2} \right)\left(1 - \frac{2Gm_r}{rc^2}\right)^{-1}. 
	\label{eqn:grel}
\end{equation}
The stars are evolved until they reach advanced stages of collapse, with infall speeds above 1,000 km s$^{-1}$. 

Masses for C1 and C2 over time are shown in Extended Data Figure~\ref{fig:cmass}.  Accretion rates in the two clumps were estimated by dividing the change in mass of each clump (as found by our halo finder) over 20 kyr intervals by this interval.  At times the rates can become negative if tidal interactions with other objects strip mass from the clump, but these outflows occur on much larger scales than accretion onto the star, which is not expected to lose mass during these episodes.  At most, accretion onto the star would temporarily halt so we set the rates to zero when the two clumps lose mass.  The rates are shown in the left panel of Extended Data Figure~\ref{fig:mdot}, where t $=$ 0 marks the onset of the collapse of the clumps.  Most of the peaks are from turbulent flows but the large dip at 1.45 Myr and subsequent spike at 1.55 Myr are due to tidal interactions that cause C2 to exchange mass with another smaller clump.  The entropy of the accreted material is matched to that of the outer layer of the star, which neglects the luminosity due to the accretion shock. However, it is at most 10$^4$ \Ls\ (\Ls, solar luminosity) even at 1 \Ms\ yr$^{-1}$ so it is small in comparison to that of the star.  We ignore stellar rotation because the spiral arms of the disks that are known to form in the clumps efficiently transport most of the angular momentum from their centers\cite{hle18a,hle18b,hle19}. There is no mass loss due to stellar winds because of low internal opacities at zero metallicity\cite{vink01,bhw01}, and we neglect pulsational mass losses because they are expected to saturate at accretion rates that are an order of magnitude below those considered here\cite{hos13}.  

Central p-p chain burning begins immediately in both stars but at very low rates that cannot halt stellar contraction, as is the case in all massive stars.  They contract until they ignite low-level triple-alpha burning at $\sim$ 750 yr and 1,400 yr in SMSs 1 and 2, respectively.  Although the triple-alpha process only provides a small contribution to the net energy generation rate, it quickly produces the carbon, nitrogen, and oxygen required to catalyze hydrogen burning via the CNO chain, which becomes the dominant energy source in SMS 1 by 9,000 yr and 10,000 yr in SMS 2.  CNO burning halts further contraction in both stars and signals the onset of main sequence burning.

The radii of the two SMSs vary from 30 -- 6,000 \Rs\ (\Rs, solar radii), or 0.14 -- 27.9 AU, over their lifetimes as seen in Extended Data Figure~\ref{fig:mdot}.  Their internal structures are shown in the Kippenhahn diagrams in Extended Data Figure~\ref{fig:kip}.  Most of the mass of the stars lies within their convective interiors, relatively little mass is in their higher-entropy outer envelopes. Modest mean accretion rates in the first Myr allow SMS 2 to become largely thermally relaxed, remaining hot and blue even with much stronger accretion after 1 Myr and becoming mostly convective after 1.2 Myr.  Both stars die when they encounter the GR instability while still on the hydrogen-burning main sequence. SMS 1 collapses with a central hydrogen fraction of just below 20\% while SMS 2 collapses earlier its evolution (after the onset of very rapid accretion at 1 Myr) with a central hydrogen fraction of $\sim$50\%.  Internal velocities for both stars are shown at the final stages of collapse in Extended Data Figure~\ref{fig:coll}.

\noindent {\bf Data Availability} \hspace{0.1in} The Enzo parameter files and initial conditions files generated by MUSIC that are required to perform the simulations are available at https://doi.org/10.5281/zenodo.5853118 (DOI: 10.5281/zenodo.5853118).  The MUSIC input files required to generate the initial conditions are available at https://sites.google.com/site/latifmaastro/ics.  The MESA inlist files required to evolve the two stars can be found at https://cococubed.com/mesa\_market/inlists.html.

\noindent {\bf Code Availability} \hspace{0.1in} The code used to produce our cosmological simulations, Enzo 2.6, can be found at https://bitbucket.org/enzo/enzo-dev/tree/enzo-2.6.1.  The MESA code used to evolve the two stars, version 12778, can be found at https://zenodo.org/record/3698354\#.YgkaN-TfUlQ.

\begin{addendum}

\item 

MAL was supported by UAEU UPAR grant No. 31S390.  SK acknowledges support from the UK STFC via grant ST/V000594/1. NPH acknowledges funding from the European Research Council for the Horizon 2020 ERC Consolidator Grant project ICYBOB, grant number 818940. TEW acknowledges support from the NRC-Canada Plaskett Fellowship.  The Enzo simulations were performed on HPC resources at UAEU and the Institute of Cosmology and Gravitation at the University of Portsmouth (Sciama).

\item[Author Contributions] DJW proposed this study, assisted in its development, and wrote the paper. MAL developed this study, performed the Enzo simulations, and analyzed its data.  SK contributed to the development of this study and the interpretation of its results.  NPH performed the MESA calculations and analyzed its results with TEW.   

\item[Competing Interests] The authors declare no competing interests.  

\item[Supplementary videos]are available in the online version of this paper.
\item[Reprints and permissions information]is available at www.nature.com/reprints.
\item[Correspondence and requests]for materials should be addressed to DJW or MAL.   

\end{addendum}

\setcounter{figure}{0}

\renewcommand{\figurename}{Extended Data Figure}
\renewcommand{\tablename}{Table~Extended Data}


\begin{figure}
\centering
\includegraphics[width=\textwidth]{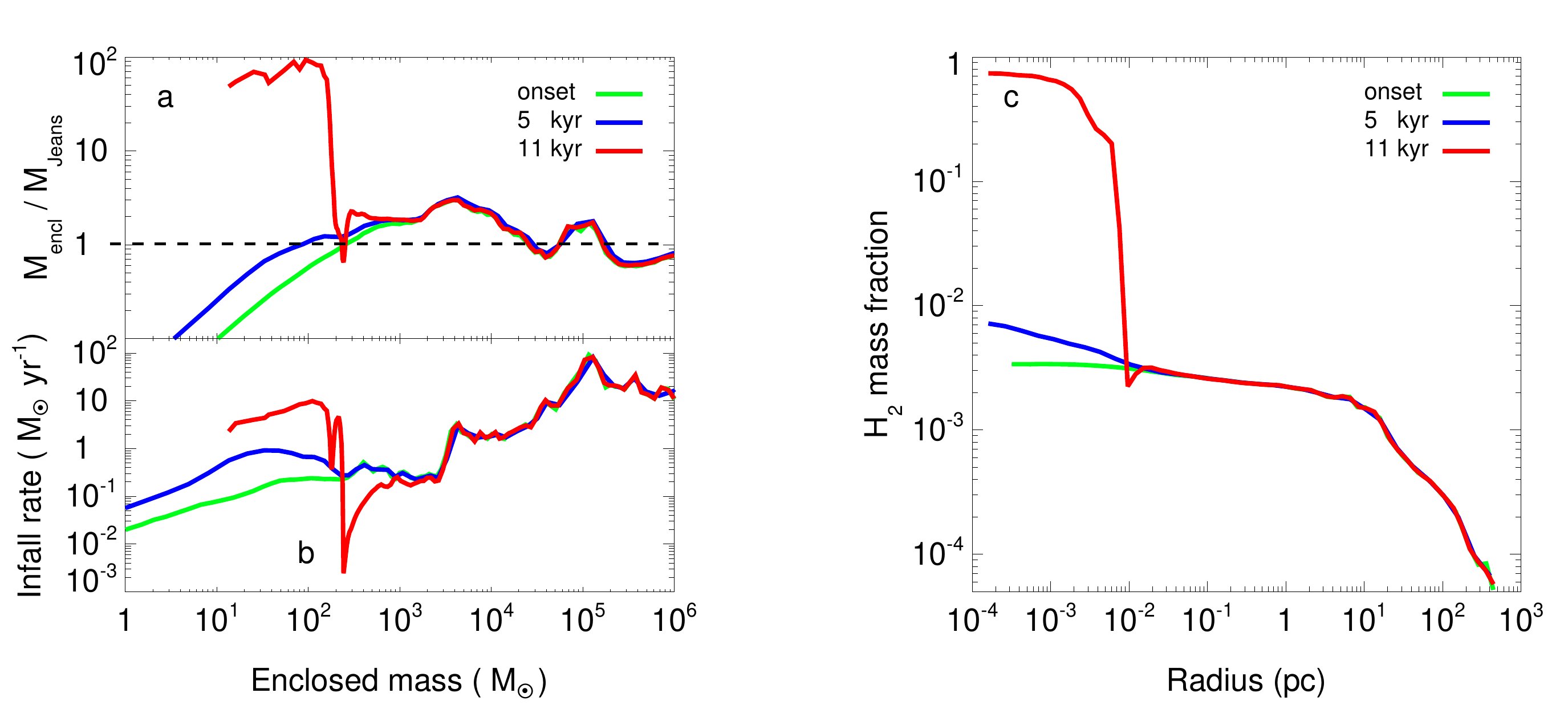} 
\caption{{\bf Additional radial gas profiles at collapse.}  The ratio of the enclosed gas mass to the Jeans mass (a) and gas infall rates (b) are at the onset of collapse (green), 5 kyr (blue), and at 11 kyr (red).  Regions where the ratio exceeds 1 are where gas is prone to collapse.  H$_2$ mass fractions (c) are at the same times.  At 11 kyr they approach unity at the smallest scales, indicating the nearly complete molecularization of the core.}
\label{fig:JM}
\end{figure}


\begin{figure}[p]
\centering
\includegraphics[width=\textwidth]{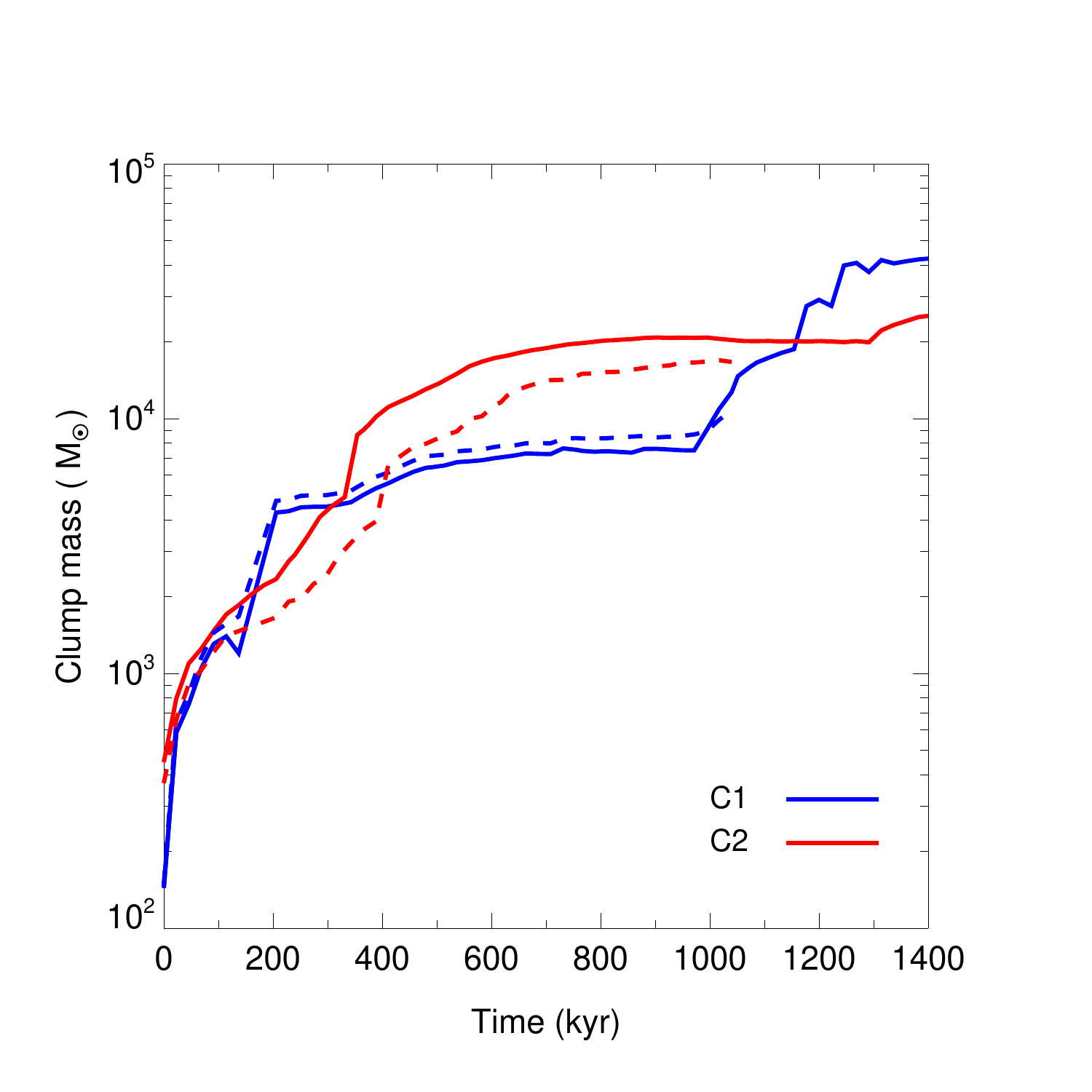}
\caption{{\bf C1 and C2 masses vs. time in the pressure floor run (solid) and the sink particle run (dashed)}.  The occasional slight dips in C1 mass are due to tidal stripping with other structures in the core.}
\label{fig:cmass}
\end{figure}


\begin{figure}[p]
\centering
\includegraphics[width=\textwidth]{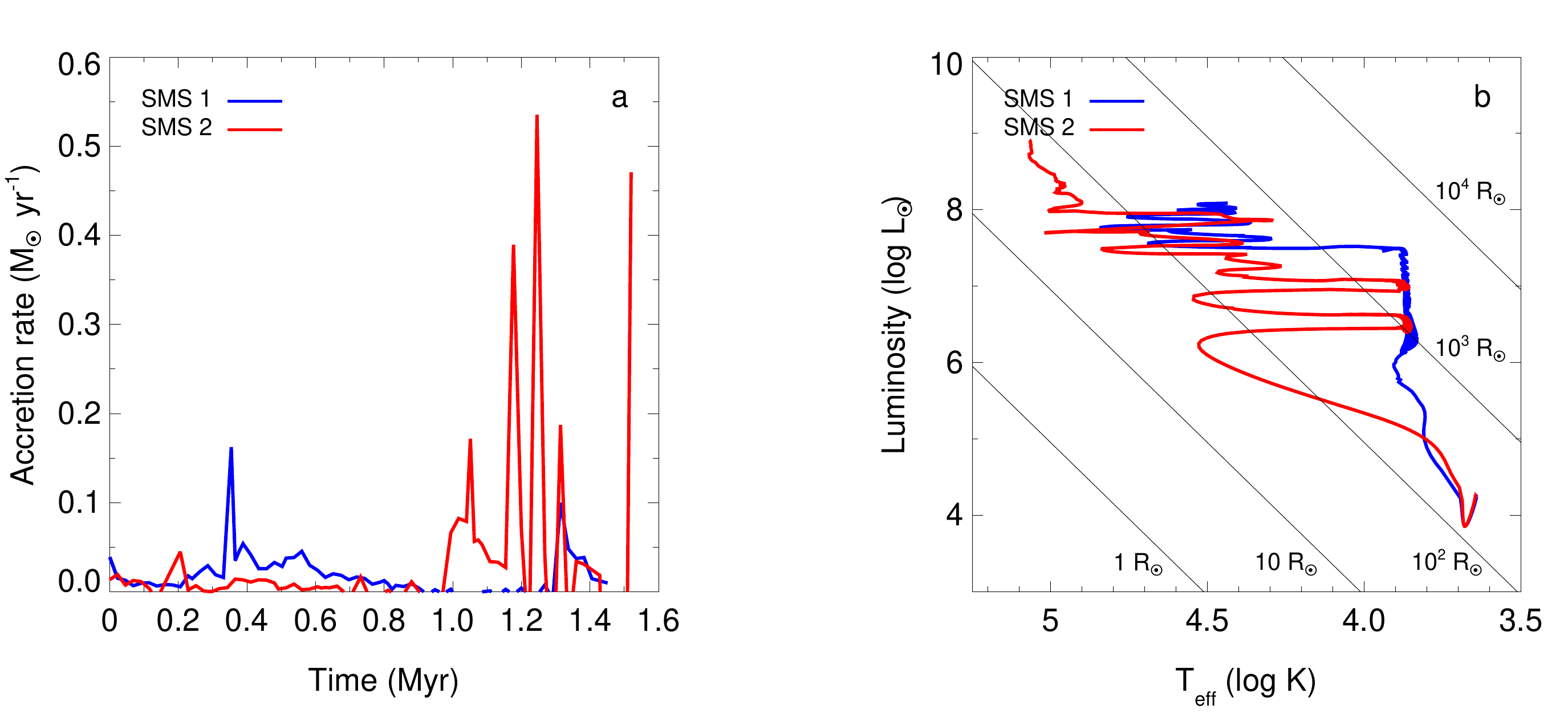} 
\caption{{\bf Evolution of SMS 1 and SMS 2.}  Accretion rates are shown in (a) and Hertzsprung-Russell tracks are shown in (b).}
\label{fig:mdot}
\end{figure}


\begin{figure}
\centering
\includegraphics[width=\textwidth]{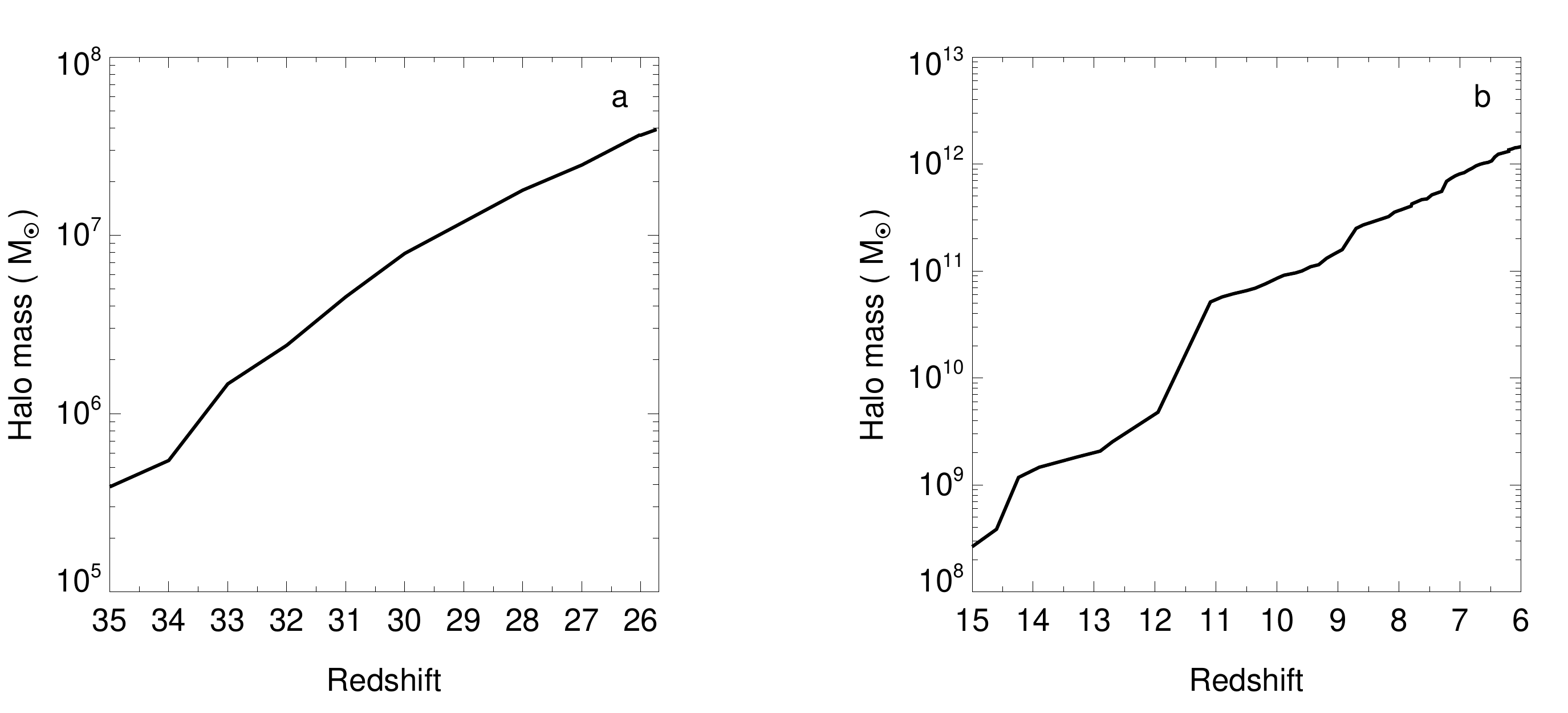} 
\caption{{\bf Halo mass vs. redshift.}  The masses in (a) are from z $=$ 35 (80 Myr after the Big Bang), when it crosses the threshold for normal Pop III star formation, to z $=$ 25.7 (126 Myr after the Big Bang), when it begins to dynamically collapse.  The subsequent growth of the halo from z $=$ 15 to z $=$ 6 is shown in (b).}
\label{fig:hmass}
\end{figure}


\begin{figure}
\centering
\includegraphics[width=\textwidth]{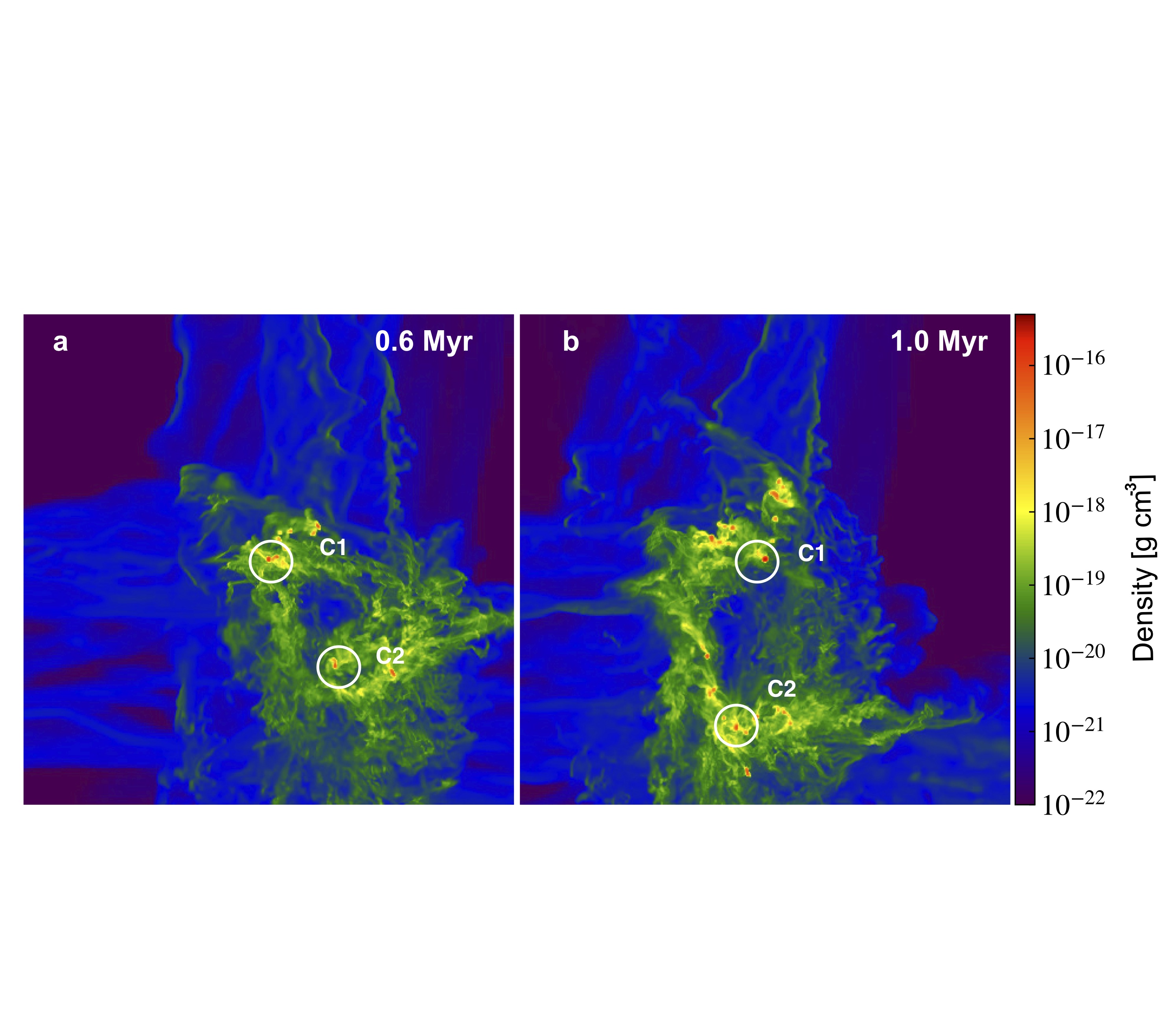}
\caption{{\bf Structure of the turbulent core in the sink particle run.}  Images are 25 pc on a side and the times in a and b are 0.6 Myr and 1.0 Myr after collapse, respectively.}
\label{fig:core2}
\end{figure}


\begin{figure}[p]
\centering
\includegraphics[width=\textwidth]{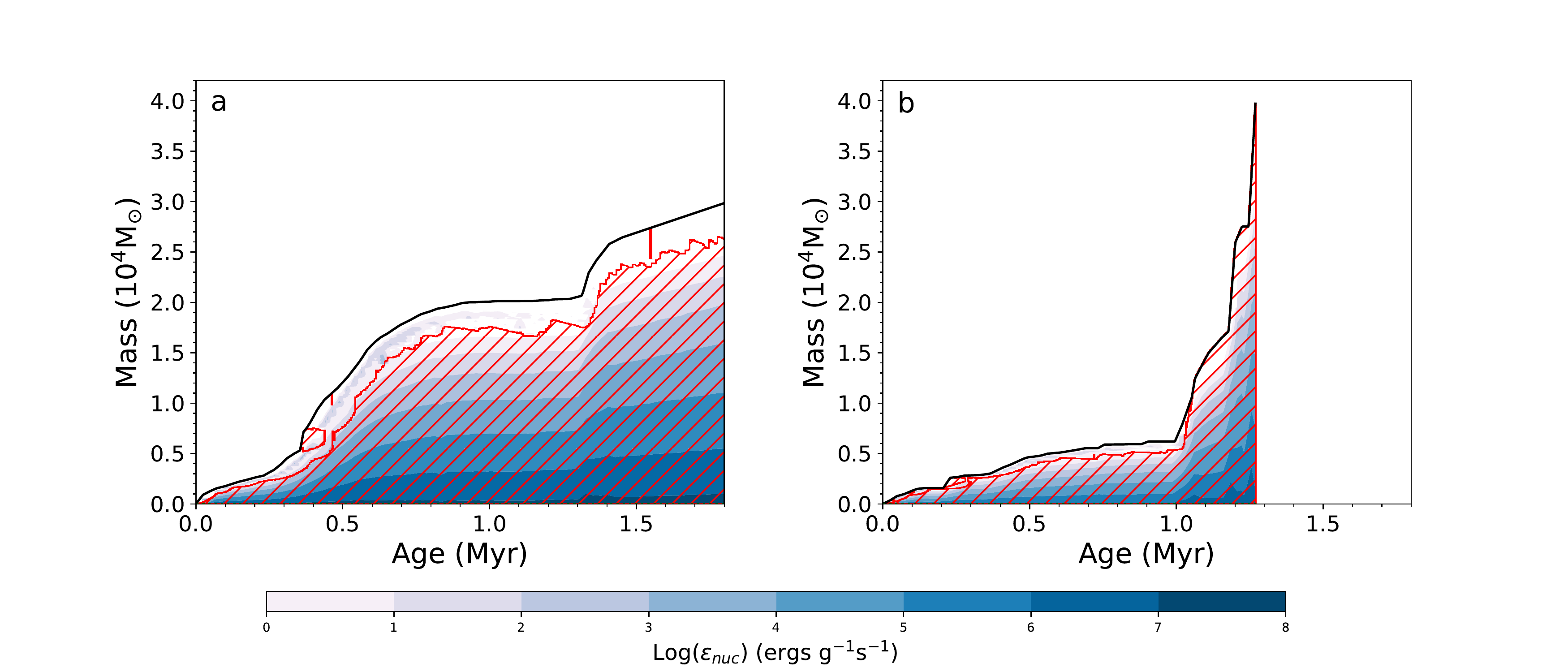} 
\caption{{\bf Kippenhahn diagrams for the two stars.}  Internal structures for SMS 1 and 2 are shown in (a) and (b), respectively.  The x- and y-axes are time and mass coordinate, respectively, so the interior structure of the star and energy generation at a given time can be read from the vertical line through the diagram at that time.  Colours show energy generation rates, ${\varepsilon}_{\mathrm{nuc}}$, at each mass coordinate and red contours mark regions where convection dominates energy transport (all other regions are radiative).}
\label{fig:kip}
\end{figure}


\begin{figure}[p]
\begin{center}
\includegraphics[width=\textwidth]{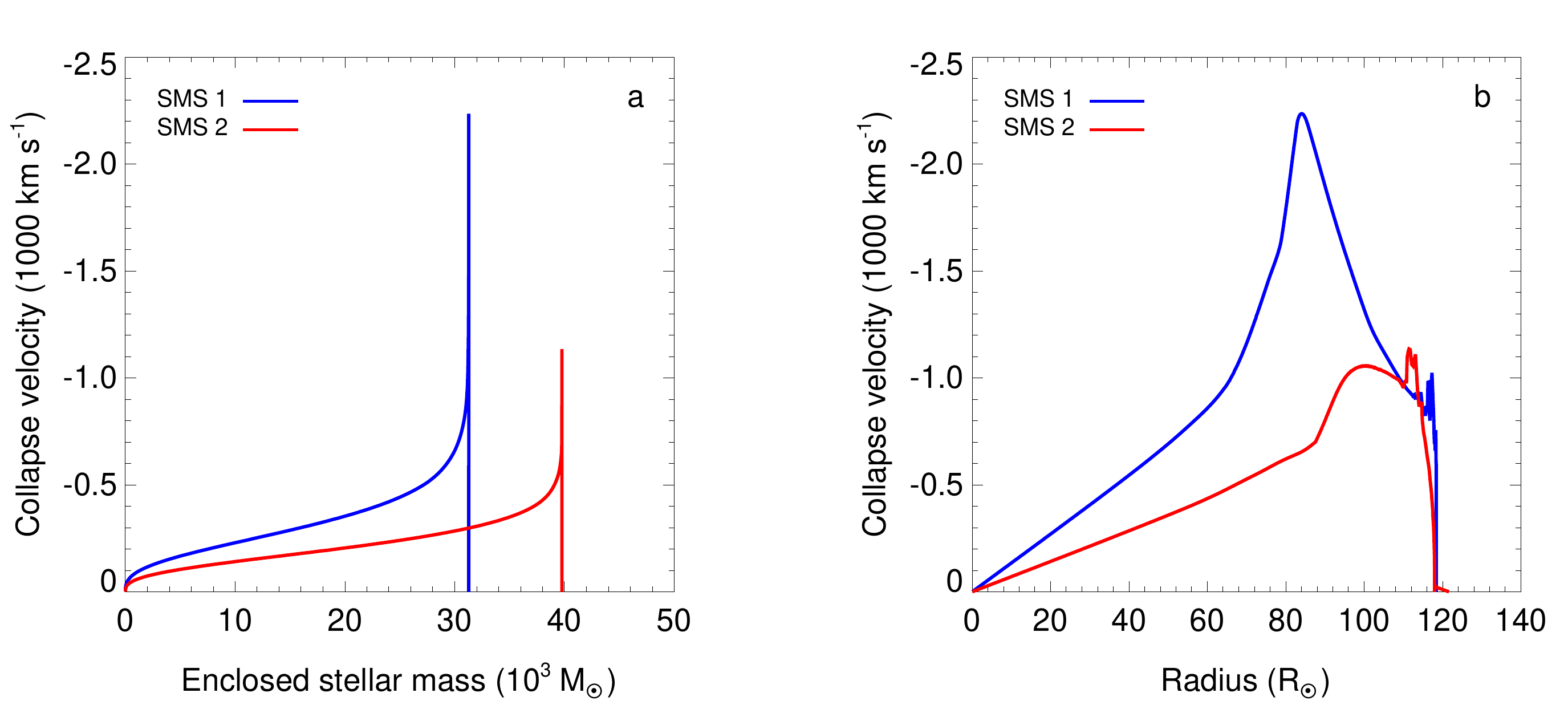}
\end{center}
\caption{{\bf The collapse of SMS 1 and SMS 2 to DCBHs.}  Infall velocities are shown vs. mass coordinate in (a) and vs. radius in (b).}
\label{fig:coll}
\end{figure}

\end{document}